\documentclass[12pt]{article}
\usepackage{latexsym}
\usepackage{amsmath}
\usepackage{amsfonts}
\usepackage{amssymb}
\usepackage[cp1250]{inputenc}

\title{A note on the Hamiltonian formalism for higher-derivative theories }
\author{K.Andrzejewski\thanks{supported by the grant 691 of University of Lodz and European Social Fund and 
Budget of State under the Integrated Regional Operational Programme}  , 
 J.Gonera\thanks{supported by the grants 690 and 795 of University of Lodz}, 
P.Ma\'slanka \thanks{supported by the grant 1 P03B 02 128 of the Polish Ministry of Science.} \\ 
Department of Theoretical Physics II, Institute of Physics,\\
University of {\L}\'od\'z ,
Pomorska 149/153, 90 - 236 {\L}\'od\'z, Poland.}

\date{}

\begin{document}
\maketitle
\begin{abstract}
An alternative version of Hamiltonian formalism for higher-derivative theories is presented. It is related
to the standard Ostrogradski approach by a canonical transformation. The advantage of the approach presented is 
that the Lagrangian is nonsingular and the Legendre transformation is performed in a straightforward way. 
\end{abstract}

\newpage
In some theories the Lagrangians containing higher time derivatives appear naturally. This concerns effective low energy theories, modified 
theories of gravity or noncommutative field theories. A standard framework for dealing with such theories on hamiltonian level is provided by 
Ostrogradski formalism \cite {b1}, \cite {b2}, \cite {b3}. The main disadvantage of the latter is that the Hamiltonian, 
being linear function of some momenta, is unbounded 
from below. In general, this cannot be cured by trying to devise an alternative cannonical formalism. In fact, any Hamiltonian is an integral of 
motion while it is by far not obvious that a generic system described by a higher-derivative Lagrangian posses globally defined integrals of 
motion except the one related to time translation invariance.

Ostrogradski approach has also some technical disadvantages. There is no straightforward transition from the Lagrangian to the Hamiltonian 
formalism. On the contrary, one has to introduce first the Lagrange multipliers enforcing the proper relations between some coordinates and 
the time derivatives of other coordinates; then the Dirac formalism of constrained dynamics is used to construct the Hamiltonian 
\cite {b2}, \cite {b3}, \cite {b4}.

Ostrogradski approach is based on the idea that the consecutive time derivatives of the initial coordinate form new coordinates, 
$q_i\sim q^{(i-1)}$. 
However, it has been suggested \cite {b5}, \cite {b6}, \cite {b7} that one can use every second derivative as a new variable, 
$q_i\sim q^{(2i-2)}$. In the present note we study 
this idea in some detail. Following Ref. \cite {b5} we modify the initial Lagrangian by adding a term which on-shell becomes a total time 
derivative. It appears that part of new coordinates can be identified with even time derivatives of the initial one. The resulting Lagrangian 
is nonsingular and the Legendre transformation can be easily performed. The Hamiltonian coincides with the Ostrogradski one while the canonical 
variables are related to the standard ones by a canonical transformation. The generating function for this transformation is given explicitly.

We start with the Lagrangian depending on time derivatives up to some even order
\begin{eqnarray}
L=L(q,\dot q,\ddot q,...,q^{(2n)})    \label{w1}
\end{eqnarray}
Define new variables
\begin{eqnarray}
&& q_i\equiv q^{(2i-2)}, \;\;\; i=1,...,n+1   \label{w2} \\
&& \dot q_i\equiv q^{(2i-1)}, \;\;\; i=1,...,n   \nonumber
\end{eqnarray}
so that 
\begin{eqnarray}
L=L(q_1,\dot q_1,q_2,\dot q_2,...,q_n,\dot q_n,q_{n+1})  \label{w3}
\end{eqnarray}
Let further $F$\ be any (at least twice differentiable) function of the following variables
\begin{eqnarray}
F=F(q_1,\dot q_1,...,q_n,\dot q_n,q_{n+1},q_{n+2},...,q_{2n})    \label{w4}
\end{eqnarray}
obeying
\begin{eqnarray}
 (i) \;\;\; \frac{\partial L}{\partial q_{n+1}}+\frac{\partial F}{\partial \dot q_n}=0 \label{w5} 
\end{eqnarray}
and
\begin{eqnarray}
 (ii) \;\;\; det\left[\frac{\partial ^2F}{\partial q_i\partial \dot q_j}\right]_{n+2\leq i\leq 2n \atop 1\leq j\leq n-1}\neq 0,\;\;\;n\geq 2   \label{w6}
\end{eqnarray}
(for $n=1$\ only $(i)$\ remains).

Finally, we define a new Lagrangian
\begin{eqnarray}
\mathcal{L}\equiv L+\sum\limits_{k=1}^n\left(\frac{\partial F}{\partial q_k}\dot q_k+\frac{\partial F}{\partial \dot q_k}q_{k+1}\right)
+\sum\limits_{j=n+1}^{2n}\frac{\partial F}{\partial q_j}\dot q_j    \label{w7}
\end{eqnarray}
Let us have a look on Lagrange equations
\begin{eqnarray}
 \frac{\partial \mathcal{L}}{\partial q_i}-\frac{d}{dt}\left(\frac{\partial \mathcal{L}}{\partial \dot q_i}\right)=0, \;\;\; i=1,...,2n  \label{w8}
\end{eqnarray}
Using (\ref {w3}), (\ref {w4}), (\ref {w7}) and (\ref {w8}) one finds 
\begin{eqnarray}
\sum\limits_{k=1}^n\frac{\partial ^2F}{\partial q_i\partial \dot q_k}(q_{k+1}-\ddot q_k)=0 \;\;\; i=n+1,...,2n   \label{w9}
\end{eqnarray}
However, eqs. (\ref {w3}), (\ref {w4}) and (\ref {w5}) imply that $\frac{\partial ^2F}{\partial q_i\partial \dot q_n}\neq 0$\ only for $i=n+1$. 
Therefore, (\ref {w6}) and (\ref {w9}) give
\begin{eqnarray}
q_{k+1}=\ddot q_k, \;\;\; k=1,...,n   \label{w10}
\end{eqnarray}
Let us now consider (\ref {w8}) for $1\leq i\leq n$. We find
\begin{eqnarray}
\frac{\partial L}{\partial q_i}-\frac{d}{dt}\left(\frac{\partial L}{\partial \dot q_i}\right)+\frac{\partial F}{\partial \dot q_{i-1}}
-\frac{d^2}{dt^2}\left(\frac{\partial F}{\partial \dot q_i}\right)=0, \;\;\; i=1,...,n    \label{w11}
\end{eqnarray}
where, by definition $\frac{\partial F}{\partial \dot q_0}=0$. By combining these equations and using (\ref {w5}) and (\ref {w10}) we arrive 
finally at the initial Lagrange equation
\begin{eqnarray}
\sum\limits_{k=0}^{2n}(-1)^k\frac{d^k}{dt^k}\left(\frac{\partial L}{\partial q^{(k)}}\right)=0  \label {w12}
\end{eqnarray}
Let us now consider the Hamiltonian formalism. Contrary to the Ostrogradski approach the Legendre transformation can be immediately performed; neither 
additional Lagrange multipliers nor constraints analysis are necessary. In fact, let us define the canonical momenta in a standard way
\begin{eqnarray}
p_i=\frac{\partial \mathcal{L}}{\partial \dot q_i}    \label {w13}
\end{eqnarray}
so that
\begin{eqnarray}
&&p_i=\frac{\partial F}{\partial q_i}, \;\;\; i=n+1,...,2n   \label {w14}\\
&& p_i=\frac{\partial L}{\partial \dot q_i}+\sum\limits_{k=1}^n\left(\frac{\partial ^2F}{\partial \dot q_i\partial q_k}\dot q_k+\frac{\partial ^2F}
{\partial \dot q_i\partial \dot q_k}q_{k+1}\right)+ \nonumber \\
&&+\sum\limits_{j=n+1}^{2n}\frac{\partial ^2F}{\partial \dot q_i\partial q_j}\dot q_j+
\frac{\partial F}{\partial q_i}, \;\;\; i=1,...,n   \label {w15}
\end{eqnarray}
By virtue of eqs.(\ref {w5}) and (\ref {w6}) eqs.(\ref {w14}) can be solved for $\dot q_1,\dot q_2,...,\dot q_n$\
\begin{eqnarray}
\dot q_i=f_i(q_1,...,q_{2n},p_{n+1},...,p_{2n}), \;\;\; i=1,...,n   \label {w16}
\end{eqnarray}
Using again eqs.(\ref {w5}) and (\ref {w6}) we can now solve (\ref {w15}) for $\dot q_i, \; n+1\leq i\leq 2n$. Finally, the Hamiltonian is 
calculated according to the standard prescription.

In order to compare the present formalism with the Ostrogradski approach let us note that they must be related by a canonical transformation. 
To see this we define new (Ostrogradski) variables $\tilde q_k,\tilde p_k, \; 1\leq k\leq 2n$:
\begin{eqnarray}
&& \tilde q_{2i-1}=q_i, \;\;\; i=1,..,n       \label {w17} \\
&& \tilde q_{2i}=f_i(q_1,...,q_{2n},p_{n+1},...,p_{2n}), \;\;\; i=1,...,n    \label {w18} \\
&& \tilde p_{2i-1}=p_i-\frac{\partial F}{\partial q_i}(q_1,f_1(...),...,q_n,f_n(...),q_{n+1},...,q_{2n}), \;\;\; i=1,...,n   \label {w19} \\
&& \tilde p_{2i}=-\frac{\partial F}{\partial f_i}(q_1,f_1(...),...,q_n,f_n(...),q_{n+1},...,q_{2n}), \;\;\; i=1,...,n    \label {w20}
\end{eqnarray}
It is easily seen that the above transformation is a canonical one, i.e. the Poisson brackets are invariant. It is not hard to find the relevant 
generating function
\begin{eqnarray}
&& \Phi (q_1,...,q_{2n},\tilde p_1,\tilde q_2,\tilde p_3,\tilde q_4,...,\tilde p_{2n-1},\tilde q_{2n})=    \label {w21} \\
&& =\sum\limits_{k=1}^nq_k\tilde p_{2k-1}+F(q_1,\tilde q_2,q_2,\tilde q_4,...,q_n,\tilde q_{2n},q_{n+1},...,q_{2n})  \nonumber
\end{eqnarray}
It is also straightforward to check that both Hamiltonians coincide. Moreover, the definitions $(\ref {w17})\div (\ref {w20})$\ reduce on-shell to 
the Ostrogradski ones.

Let us now consider the case of Lagrangian depending on time derivatives up to some odd order
\begin{eqnarray}
L=L(q,\dot q,\ddot q,...,q^{(2n+1)})    \label {w22}
\end{eqnarray}
Again, we define
\begin{eqnarray}
&& q_i\equiv q^{(2i-2)}, \;\;\; i=1,...,n+1   \label {w23}  \\
&& \dot q_i\equiv q^{(2i-1)},  \;\;\; i=1,...,n+1   \label {w24}
\end{eqnarray}
so that
\begin{eqnarray}
L=L(q_1,\dot q_1,q_2,\dot q_2,...,q_{n+1},\dot q_{n+1})     \label {w25}
\end{eqnarray}
Now, let us select a function $F$,
\begin{eqnarray}
F=F(q_1,\dot q_1,q_2,\dot q_2,...,q_n,\dot q_n,q_{n+1},...,q_{2n+1})    \label {w26}
\end{eqnarray}
subject to the single condition
\begin{eqnarray}
det\left[\frac{\partial ^2F}{\partial q_i\partial \dot q_k}\right]_{n+2\leq i\leq 2n+1\atop 1\leq k\leq n}\neq 0   \label {w27}
\end{eqnarray}
and define the Lagrangian
\begin{eqnarray}
\mathcal{L}=L+\sum\limits_{k=1}^n\left(\frac{\partial F}{\partial q_k}\dot q_k+\frac{\partial F}{\partial \dot q_k}q_{k+1}\right)+
\sum\limits_{j=n+1}^{2n+1}\frac{\partial F}{\partial q_j}\dot q_j  \label {w28}
\end{eqnarray}
Consider the Lagrange equations (\ref {w8}). First, we have
\begin{eqnarray}
\sum\limits_{k=1}^n\frac{\partial ^2F}{\partial q_i\partial \dot q_k}(q_{k+1}-\ddot q_k)=0, \;\;\; i=n+2,...,2n+1   \label {w29}
\end{eqnarray}
and, by virtue of (\ref {w27})
\begin{eqnarray}
q_{k+1}=\ddot q_k, \;\;\; k=1,...,n   \label {w30}
\end{eqnarray}
The remaining equations read
\begin{eqnarray}
\frac{\partial L}{\partial q_i}-\frac{d}{dt}\left(\frac{\partial L}{\partial \dot q_i}\right)+\frac{\partial F}{\partial \dot q_{i-1}}-
\frac{d^2}{dt^2}\left(\frac{\partial F}{\partial \dot q_i}\right)=0, \;\;\; i=1,...,n+1    \label {w31}
\end{eqnarray}
with $\frac{\partial F}{\partial \dot q_0}=0, \; \frac{\partial F}{\partial \dot q_{n+1}}=0$. Combining (\ref {w30}) and (\ref {w31}) one gets
\begin{eqnarray}
\sum\limits_{k=0 }^{2n+1}(-1)^k\frac{d^k}{dt^k}\left(\frac{\partial L}{\partial q^{(k)}}\right)=0    \label {w32}
\end{eqnarray}
Let us note that no condition of the form (\ref {w5}) is here necessary.

Also in the odd case the present formalism is related to that of Ostrogradski by a canonical transformation. Indeed, the canonical momenta read 
\begin{eqnarray}
&& p_i=\frac{\partial F}{\partial q_i}, \;\;\; i=n+2,...,2n+1     \label {w33}  \\
&& p_i=\frac{\partial L}{\partial \dot q_i}+\sum\limits_{k=1}^n\left(\frac{\partial ^2F}{\partial \dot q_i\partial q_k}\dot q_k+
\frac{\partial ^2F}{\partial \dot q_i\partial \dot q_k}q_{k+1}\right)+  \nonumber \\
&& +\sum\limits_{j=n+1}^{2n+1}\frac{\partial ^2F}{\partial \dot q_i\partial q_j}
\dot q_j,  \;\;\; i=1,...,n+1   \nonumber
\end{eqnarray}
and, as previously, these equations can be solved in terms of velocities, in particular
\begin{eqnarray}
\dot q_i=f_i(q_1,...,q_{2n+1},p_{n+2},...,p_{2n+1}), \;\;\; i=1,...,n    \label {w34}
\end{eqnarray}
Now, one can define the canonical transformation to Ostrogradski variables
\begin{eqnarray}
&& \tilde q_{2i-1}=q_i, \;\;\; i=1,...,n+1   \label {w35}   \\
&& \tilde q_{2i}=f_i(q_1,...,q_{2n+1},p_{n+2},...,p_{2n+1}), \;\;\; i=1,...,n    \label {w36}  \\
&& \tilde p_{2i-1}=p_i-\frac{\partial F}{\partial q_i}(q_1,f_1(...),...,q_n,f_n(...),q_{n+1},...,q_{2n+1}), \;\;\; i=1,...,n+1    \label {w37} \\
&& \tilde p_{2i}=-\frac{\partial F}{\partial f_i}(q_1,f_1(...),...,q_n,f_n(...),q_{n+1},...,q_{2n+1}), \;\;\; i=1,...,n   \label {w38}
\end{eqnarray}
The relevant generating function reads
\begin{eqnarray}
&& \Phi (q_1,q_2,...,q_{2n+1},\tilde p_1,\tilde q_2,\tilde p_3,\tilde q_4,...,\tilde q_{2n},\tilde p_{2n+1})=   \nonumber \\
&& =\sum\limits_{k=1}^{n+1}\tilde p_{2k-1}q_k+F(q_1,\tilde q_2,...,q_n,\tilde q_{2n},q_{n+1},...,q_{2n+1})    \label {w39}
\end{eqnarray}
Sumarizing, we have found a modified Lagrangian and Hamiltonian formulations of higher-derivative theories. They are equivalent to the Ostrogradski 
formalism in the sense that on the Hamiltonian level they are related by a canonical transformation. However, the advantage of the approach presented 
is that the Lagrangian is nonsingular and the Legendre transformation can be performed in a straightforward way. 

Let us conclude with the simple example. Consider the Lagrangian \cite{b6}, \cite{b8}, \cite{b9}
\begin{eqnarray}
L=\frac{1}{2}\dot q^2-\frac{\omega ^2}{2}q^2-gq\ddot q   \label{w40}
\end{eqnarray}
and define
\begin{eqnarray}
&& q_1=q, \;\;\; q_2=\ddot q   \label{w41}  \\
&& \mathcal{L}=\frac{1}{2}\dot q_1^2-\frac{\omega ^2}{2}q_1^2-gq_1q_2^2+\frac{\partial F}{\partial q_1}\dot q_1+\frac{\partial F}{\partial \dot q_1}q_2
+\frac{\partial F}{\partial q_2}\dot q_2   \nonumber
\end{eqnarray}
with $F$\ obeying
\begin{eqnarray}
\frac{\partial F}{\partial \dot q_1}-2gq_1q_2=0  \label{w42}
\end{eqnarray}
Eq.(\ref{w42}) implies
\begin{eqnarray}
F(q_1,\dot q_1,q_2)=2gq_1\dot q_1q_2+W(q_1,q_2)    \label{w43}
\end{eqnarray}
By neglecting total time derivative one obtains by virtue of eqs.(\ref{w41}), (\ref{w43})
\begin{eqnarray}
\mathcal{L}=\frac{1}{2}\dot q_1^2-\frac{\omega ^2}{2}q_1^2+gq_1q_2^2+2g\dot q_1^2q_2+2gq_1\dot q_1\dot q_2    \label{w44}
\end{eqnarray}
It is now straightforward to construct the relevant Hamiltonian
\begin{eqnarray}
H=\frac{p_1p_2}{2gq_1}-\frac{(1+4gq_2)}{8g^2q_1^2}p_2^2+\frac{\omega ^2}{2}q_1^2-gq_1q_2^2   \label{w45}
\end{eqnarray}
and one easily checks that the corresponding canonical equations yield the initial equation for $q\equiv q_1$.

The generating function for canonical transformation to Ostrogradski variables reads
\begin{eqnarray}
\Phi (q_1,q_2,\tilde p_1,\tilde q_2)=q_1\tilde p_1+2gq_1q_2\tilde q_2   \label{w46}
\end{eqnarray}
and gives
\begin{eqnarray}
&& q_1=\tilde q_1   \nonumber \\
&& q_2=-\frac{\tilde p_2}{2g\tilde q_1}   \label{w47} \\
&& p_1=\tilde p_1-\frac{\tilde p_2\tilde q_2}{\tilde q_1}   \nonumber  \\
&& p_2=2g\tilde q_1\tilde q_2   \nonumber
\end{eqnarray}
In terms of new variables the Hamiltonian (\ref{w45}) reads
\begin{eqnarray}
H=\tilde p_1\tilde q_2-\frac{\tilde p_2^2}{4g\tilde q_1}-\frac{1}{2}\tilde q_2^2+\frac{\omega ^2}{2}\tilde q_1^2   \label{w48}
\end{eqnarray}
and coincides with the Ostrogradski Hamiltonian.

Let us note that both Hamiltonians (\ref{w45}) and (\ref{w48}) are singular for $g=0$\ and $q_1(=\tilde q_1)=0$. 
This is due to the fact that in this case the Lagrangian (\ref{w40}) reduces from second to first order in time derivatives.

\vspace{12pt}

{\large\bf Acknowledgement}
Thanks are due to Prof. Piotr Kosi\'nski and Prof. Michal Majewski for interesting discussions and useful remarks.

\end{document}